\documentstyle[emulateapj,psfig]{article}

\begin{document}

\title{Composite Spectra from the {\em FIRST} Bright Quasar Survey}
\author{M. S. Brotherton\altaffilmark{1,2}, Hien D. Tran\altaffilmark{2,3}, 
R. H. Becker\altaffilmark{2,4}, Michael D. Gregg\altaffilmark{2,4}, \\
S. A. Laurent-Muehleisen\altaffilmark{2,4}, R. L. White\altaffilmark{5}}

\altaffiltext{1}{KPNO/NOAO, 950 N. Cherry Ave., P.\ O.\ Box 26732, Tucson,
AZ 85726}
\altaffiltext{2}{Institute of Geophysics and Planetary Physics, Lawrence Livermore National Laboratory, 7000 East Avenue, P.O. Box 808, L413, Livermore, CA 94550}
\altaffiltext{3}{Current Address: Johns Hopkins University, Baltimore, MD 21218}
\altaffiltext{4}{Physics Dept., University of California--Davis}
\altaffiltext{5}{Space Telescope Science Institute, 3700 San Martin Dr., 
Baltimore, MD 21218}

\begin{abstract}

We present a very high signal-to-noise ratio composite spectrum created
using 657 radio-selected quasars from the FIRST Bright Quasar Survey.
The spectrum spans rest-frame wavelengths 900 -- 7500 \AA.  Additionally
we present composite spectra formed from subsets of the total data set in
order to investigate the spectral dependence on radio loudness and 
the presence of broad absorption.  In particular, radio-loud quasars are red
compared to radio-quiet quasars, and quasars showing low-ionization
broad absorption lines are red compared to other quasars.
We compare our composites with those from the Large Bright Quasar Survey.  
Composite quasar spectra
have proven to be valuable tools for a host of applications, and in that
spirit we make these publically available via the FIRST survey web page.

\end{abstract}
\keywords{quasars: emission lines, quasars: general}
\section{Introduction}

For over three decades, the construction of composite spectra has proven to 
be a valuable activity in AGN research.  The synthesis of broad-band colors
of a large number of quasars produced low-resolution ``compromise composite'' 
spectra, enabling the average quasar continuum to be studied over a large 
wavelength range and $k$-corrections to be computed (Sandage 1966; Evans \& Hart
1977).  Emission-line intensity ratios for a large number of quasars over a 
range of redshifts were combined to determine a composite quasar ``spectrum''
(Chan \& Burbidge 1975), and it was this technique that first identified
the low L$\alpha$/H$\beta$ ratio that posed a problem for photoionization
models (Baldwin 1977).  A similar bootstrap method established the average
emission-line velocity shifts in quasars (Tytler \& Fan 1992).

Technological advances led to large samples of quasars with 
readily manipulated digital data, resulting in myriad composite spectra in the
1990s.  These included composite quasar spectra presented by Boyle (1990), 
Cristiani \& Vio (1990), and Francis et al. (1991 -- using the LBQS, i.e. the 
Large Bright Quasar Survey), resulting in 
identification of many weak emission lines and additional recognition of
the significance of the omnipresence of Fe II emission blends throughout
the ultraviolet and optical.  
Zheng et al. (1997) produced a composite spectrum using 
quasar spectra from the {\em Hubble Space Telescope} with unprecedented 
coverage of short wavelengths.  These composites have been used extensively 
for many applications.

Composite spectra also permit investigation of the spectral dependencies
on other properties.  Sprayberry \& Foltz (1992)
compared the composite spectra of quasars with broad absorption lines (BALs)
to those without, finding that quasars with low-ionization BALs 
have significantly redder continua consistent with line-of-sight dust.
Francis, Hooper, \& Impey (1993), using the LBQS data set, created composite 
spectra of radio-loud and radio-quiet quasars separately, finding 
an extra low-velocity emission line component in the former 
(also see Cristiani \& Vio 1990 and Zheng et al. 1997).  
Brotherton et al. (1994)
created composites from two samples of quasars distinguished by the 
velocity width of the broad lines, finding significant differences in the
line ratios of the two.  Green (1998) created two composites distinguished by
X-ray brightness, finding in particular X-ray bright quasars had 
significantly stronger narrow-line emission.
Baker \& Hunstead (1995) created composites using samples distinguished by
their radio structure, finding that radio-loud quasars that appear to be
more edge-on also appear to be dustier.
Malhotra (1997) created a composite spectrum by aligning the redshifts of 
intervening Mg II absorbers present in quasar spectra and detected the 
signature of the 2175 \AA\ feature associated with dust from the absorbers.

In this paper, we present a very high-quality composite spectrum constructed 
using 657 quasars from the FIRST Bright Quasar Survey (FBQS) 
(White et al. 2000), 
as well as several composite spectra formed using subsamples of the FBQS.  
Section 2 describes the sample selection and data set used to
construct the composite spectra, as well as the algorithm we use to construct
composite spectra.  Section 3 describes the composite spectra themselves.  
Section 4 discusses the properties of the FBQS we can discern from the 
composite spectra and compares our results with those of 
the Large Bright Quasar Survey (hereafter LBQS)
(Foltz et al. 1987, 1989; Hewett et al. 1991; Chaffee et al. 1991; Morris et al.
1991; Francis et al. 1991; Francis et al. 1992). 
Section 5 summarizes our results.

\section{Sample and Data}

The FBQS starts with radio sources found using the 
NRAO\footnote{The National Radio Astronomy Observatory is a facility of
the National Science Foundation operated under cooperative agreement by
Associated Universities, Inc..} 
Very Large Array $FIRST$\ survey (Becker et al. 1995), a 
20 cm survey at $\sim 5\arcsec $\ resolution with a detection limit
$\sim$ 1 mJy.  Quasar candidates are determined by matching FIRST sources
within 1$\farcs$2 of stellar sources with $O-E < 2.0$ and $E<17.8$ from the 
Automated Plate Measuring Facility (APM) catalog of the Palomar 
Observatory Sky Survey I (POSS-I) plates (MacMahon et al. 2000).
The choice of selection criteria yields quasars and BL Lac objects with
60\% selection efficiency and misses very few bright red quasars
(Gregg et al. 1996 -- the pilot investigation).  
White et al. (2000) presents the second installment of the FBQS
with a catalog of 636 quasars distributed over 2682 deg$^2$,
fully characterizing the sample and displaying nearly all the spectra we 
use to construct our composites.  The spectra are a heterogenous set,
coming from a half dozen telescopes, although generally have 
$\sim 10$ \AA\ resolution and signal-to-noise ratios of 20 or higher.
Both radio-loud and radio-quiet quasars are 
present in approximately equal numbers, as are radio-loud BAL quasars 
(Becker et al. 2000), and large numbers of radio-intermediate quasars.
This is the first radio-selected sample that is competitive in size with 
optically selected quasar surveys such as the LBQS. 

\section{Composite Spectra}

\subsection{Methodology and Caveats}

In constructing our composite spectra, we use essentially the same
algorithm described by Francis et al. (1991) which they used to construct
their LBQS composite quasar spectrum.  In fact, we employ the same code
that has been used to construct the composite spectrum of the total
LBQS sample (S. Morris 1999, private communication), with which we make
a comparison in \S\ 4.  Below we describe the procedure, which employs IRAF
except as noted.

The spectra are rebinned to the same wavelength range,
3200 \AA\ -- 9200 \AA.  Zeroes are used for regions within this range with
no data (these zero-padded regions are given zero weight in the combination).  
These spectra are then stacked into a single image.  A similar procedure is 
followed to produce a stacked image of noise spectra.

Noise spectra are not available for all the spectra in the FBQS database
and we have used a single uniform method to estimate the noise for all
spectra which will be used to weight individual spectra in the composite.  
A high-order continuum fit is used to normalize each spectrum.  
We then use a VISTA script to measure the average S/N from these.  The spectra
are smoothed by 9 pixels, their square root is taken, then they are 
multiplied by the average S/N which provides an estimate of the 
true S/N spectrum.  The S/N is set to 1 across the atmospheric A and
B bands to ensure these regions receive low weight in the combination.
The spectra are weighted by (S/N)$^2$ when combined.

A separate file is also prepared that contains the redshifts of the 
input spectra, determined by using cross-correlation against a preliminary
composite spectrum and refined in an iterative manner.
The final composite spectrum clearly resolves the [O III] 
$\lambda\lambda$4959,5007 narrow lines (each with a FWHM of 750 km s$^{-1}$), 
while initial versions of the composite using visually assigned redshifts 
showed broader lines that were somewhat blended.
Our success with using cross-correlation redshifts mirrors the 
experience of Francis et al. (1991) in constructing the LBQS composite.

The flux and noise spectral images are stacked into a single 
three-dimensional data cube, which is the input for the IRAF procedure
provided by Simon Morris.  The procedure uses the algorithm described by
Francis et al. (1991).  Spectra are added, from low-$z$ to high-$z$, 
to a running average.  All spectra are rebinned to a common dispersion.  
The squares of the signal-to-noise ratios are used to weight the individual 
spectra.  New spectra are normalized to the running average in the 
region of overlap only.  A standard deviation is computed.  

As discussed by Francis et al. (1991), such a procedure has a number of
limitations and uncertainties.  All of the caveats that applied there also
apply to the FBQS composite.  Galactic and intrinsic extinction may 
artificially redden individual spectra, although the FBQS is selected from
objects at high Galactic latitude and a (mild) color criterion is imposed.
Slit losses from differential refraction may redden individual spectra,
although the majority of spectra were obtained at Lick and Keck Observatories 
where the spectrograph slits were always at the parallactic angle.  The
appearance of features in the composite spanning large wavelengths 
(e.g., the overall continuum shape and the 3000 \AA\ ``little blue bump'') 
is sensitive to the order spectra are added to the composite;
in the LBQS, Francis et al. (1991) reported that the spectral index 
$\alpha$ varied by $\pm$0.2, which is consistent with our own tests we
discuss in \S\ 3.2.  In a flux-limited sample such as the FBQS or LBQS,
quasars at low $z$ are typically lower luminosity than those found at
high $z$ (see Francis et al. 1991 and White et al. 2000 for illustrative 
plots), and spectral dependencies on luminosity (e.g., Korista et al. 1998)
may lead to spurious results in comparing optical to ultraviolet properties.
Finally, the arithmetic mean of power-law spectra is
not necessarily a power-law spectrum with the mean power-law index.

Below we characterize the total FBQS composite spectrum quantitatively in some
detail.  We provide less detail for the composites formed from the 
subsamples based on radio and absorption properties; because of the 
above caveats the results of our comparisons should be regarded as suggestive
but not conclusive without follow-up investigations.  These composite 
spectra are useful tools and a powerful way to graphically illustrate
certain dependencies, but the reader is cautioned not to 
overinterpret these results. 

\subsection{FBQS Composite Quasar Spectrum}

About 90\% of the 657 spectra used in construction of the FBQS composite 
are plotted by White et al. (2000) and additional details may be found there.
Confirmed and probable BAL quasars have been excluded from the composite.
Figure 1 plots the 
composite spectrum and labels major emission lines.  Figure 2 plots the
histogram of the number of quasars contributing at each wavelength.
Figure 3 plots the standard deviation spectrum; it differs from the mean
spectrum most noticeably in the strength of the narrow lines, particularly
[O III] $\lambda\lambda$4959,5007.  The signal-to-noise ratio
per 0.6 \AA\ pixel peaks at just under 200 (estimated by using the standard
deviation spectrum and the number of objects contributing to each pixel to
calculate the standard error in the mean); rebinning to 2.5 \AA\
increases the signal-to-noise ratio to just under 400.
Figure 4 plots the composite in log-log space in which a power-law is
represented as a straight line; a power-law with a spectral index $\alpha$
=$-$0.46 (F$_{\nu} \propto \nu^{\alpha}$) appears to be a good estimate of 
an underlying continuum between H$\beta$ and L$\alpha$.

Table 1 lists emission features present in the spectrum along with 
measurements of line ratios and equivalent widths.  The choices of 
integration windows are made to match those of Francis et al. (1991) 
and to facilitate comparison to the LBQS composite spectrum (\S\ 4).  
In making measurements of the standard deviation of equivalent widths it is 
assumed that the continuum and line flux vary independently.

In order to investigate to what extent the spectral energy distribution may
be a function of the normalization process, we also constructed a composite
beginning with the largest redshift quasars and working backwards to the
lowest redshift quasars, the reverse of the procedure used to construct the
spectrum in Figure 1.  The result was a very similar but slightly redder 
spectrum, also well fit by a power-law but with a spectral index $\alpha$
=$-$0.49 (which gives rise to a 12\% difference at L$\alpha$ vs.\ H$\beta$);
the emission-lines are identical between the two and divide out as expected.

\subsection{Composites with Different Radio-Loudness}

We can categorize the FBQS quasars according to their radio-loudness.
As recommended by Weymann (1997), we adopt log R* as the formal measure
of radio-loudness with log R* = 1.0 dividing radio-loud and radio-quiet
subsamples, where R* is the $K$-corrected ratio of radio-to-optical power
(Sramek \& Weedman 1980; Stocke et al. 1992).  Log R* is tabulated for
the FBQS in White et al. (2000).  
The FIRST quasars, while radio-selected, in fact include a large fraction 
of radio-quiet quasars.
Using the R* criterion, our data set contains
59\% radio-loud quasars with redshifts up to $z=3.4$ and 41\% radio-quiet 
quasars with redshifts up to $z=3.3$ (see Fig. 14 of White et al. 2000).  
The range in optical luminosities is also quite similar, although 
there exist significant differences in the luminosity ranges compared
at low $z$\ vs.\ high $z$ that result from the FBQS being a magnitude-limited
sample. Figure 5 plots the radio-loud and radio-quiet quasar composite spectra.

Compared to the radio-quiet composite, the radio-loud spectrum has a
redder spectral energy distribution, broader Balmer lines, stronger [O
III] emission, and a stronger red wing/weaker blue wing asymmetry to
the C IV $\lambda$1549 emission line profile. 
Similar differences have been previously noted (e.g., Boroson \& Green 1992;
Barthel, Tytler, \& Thomson 1990).

We also experimented with three divisions of radio-loudness, such that
radio-intermediate quasars were distinguished as those with
$0.5 <$\  log R* $< 1.5$.  
The FBQS is sensitive to radio-intermediate quasars, which are notably 
missing from earlier searches (White et al. 2000).  
The distribution of ``radio-loudness'' is not significantly bimodal.
The radio-intermediate quasar composite spectrum closely resembles
that of the radio-loud quasars, and so differs from the radio-quiet
quasar composite spectrum in a very similar manner.

\subsection{BAL Quasar Composites}

We have constructed composite spectra of samples of broad absorption line
(BAL) quasars in the FBQS. These samples include high-ionization BAL quasars 
(25 objects) and the low-ionization BAL quasars 
(18 objects, including 4 metastable Fe absorbed quasars).
The majority of objects are cataloged and characterized by Becker et al. (2000)
(29 objects, 15 high-ionization BAL quasars and 14 low-ionization BAL quasars);
in order to increase our sample sizes and the significance of the results,
we used all the BAL quasars in the FBQS database whether published or not.
Note that we have erred on the side of inclusivity, as discussed by 
Becker et al. (2000), and included several confirmed and probable BAL quasars 
that do not meet the criterion of positive BALnicity index (Weymann et al. 
1991), e.g. FIRST J1603+3002, which has been shown to
have an intrinsic high-velocity outflow (Arav et al. 1999).  
Figure 6 plots these composite spectra as well as the total FBQS composite.

The colors of the low-ionization BAL quasar composite are redder than
those of the high-ionization BAL quasar composite, which are in turn 
redder than that of the FBQS composite.  The continuum band colors of
Yamamoto \& Vansevicius (1999) indicate that the BAL quasar composites
are consistent with reddening the FBQS composite according to a
Small Magallenic Cloud (SMC) extinction law.
If we assume such a law, the spectral shape of the high-ionization BAL
quasar composite is consistent with the FBQS composite if dereddened by
$E(B-V) \sim 0.04$ mag, while the low-ionization BAL quasar composite 
must be dereddened by $E(B-V) \sim 0.1$ mag.  Small color differences
in the rest-frame optical such as these become magnified in the rest-frame
ultraviolet as seen in the figure.

BALs are not so readily apparent in the high-ionization 
BAL quasar composite spectrum.  This results from a wide range in BAL
properties, primarily the velocity offsets from the systemic redshift,
coupled with a paucity of quasars with extremely deep and broad troughs 
(e.g., as seen in PHL 5200).
The BAL features are therefore diminished in the averaging process.

\section{Discussion}

We restrict much of our discussion to a comparison between our FBQS 
composite spectra and similar ones created from the LBQS data set.
This is appropriate because of the similarities between the surveys
in terms of luminosities, redshift range, and size.  This is also 
appropriate as composites from the LBQS have been extensively used 
for a decade, and include composites made from subsamples with 
BALs and with differing degrees of radio loudness.  Finally, we can
compare our FBQS composite quasar spectrum to a LBQS composite quasar
spectrum using exactly the same algorithm and software.
Table 2 summarizes the FBQS and LBQS samples explored here with 
composite spectra, and a few of the properties of the composite spectra.

The LBQS composite spectrum of Francis et al. (1991) was created using
688 of the total LBQS sample of 1018 quasars, or 68\%.  
Following the completion of the
LBQS a composite spectrum was created using the total data set and
has been made available to us along with the software used to create it
(S. Morris 1999, private communication).  
The properties of the last $\sim$ 300 quasars are consistent with those
which went into the first composite, but the two composites 
differ significantly in a number of ways.  The source of the differences
is not entirely clear although it has been speculated that
the difference has to do with the vagaries of the normalization
procedure (P. Francis 1999, private communication).

Figure 7 compares the LBQS composite quasar spectra of Francis et al. (1991)
and Morris (1999, private communication).  
It is readily apparent that the shape of 
the spectral energy distributions differ, with 10\% excursions in the optical 
and near-UV, to a 20\% difference in the far-UV
(keeping in mind the presence of the Lyman $\alpha$ forest that 
depresses the intrinsic quasar continuum at these wavelengths).
The original LBQS survey spectra suffered from differential atmospheric slit 
losses, and the degree of corrections made or not made to the spectra at 
the time of the construction of the composite spectra likely contribute 
to the changes in the large scale continuum shape 
(P. Francis 2000, private communication).

The ultraviolet emission lines also differ between the LBQS composite spectra.
Notably, the L$\alpha$, N V blend is
25\% stronger in the new composite, and the C IV $\lambda$1549 line is
10\% stronger and may have a rather different profile.  These differences
should be kept in mind when making comparisons to tabulated line ratios
and equivalent widths in Francis et al. (1991).  The weakness of the 
L$\alpha$ and C IV $\lambda$1549 lines in the Francis et al. (1991)
composite may result from the inclusion of objects with strong associated  
absorption or weak BALs, which are probably appropriate to include for 
the purpose of making $k$-corrections but compromise measuring accurate
intrinsic average line ratios.

In comparing the FBQS composite quasar spectrum to that of the LBQS,
the different biases in sample selection and the ensuing sample differences
should first be made clear.  The LBQS used purely optical techniques to
select candidate quasars, resulting in a very different distribution of
radio-loudness in that sample (approximately 10\% radio-loud quasars --
see Hooper et al. 1995).  
While about half the FBQS quasars are formally radio-quiet, the more
extreme radio-quiets and all radio-silent quasars are not represented,
especially at high redshift.
The magnitude range of the LBQS 
($16.0 \leq m_{B_J} \leq 18.85$) is different from that of the FBQS
($E \leq 17.8$).  The inclusion of very bright sources in the FBQS may explain
why its distribution of redshifts includes a larger fraction of low-$z$
objects compared to the LBQS; the redshift ranges are similar.    

The overall spectral shape of the FBQS composite spectrum is more similar
to that of the total LBQS composite spectrum than that of Francis et al.
(1991).  The FBQS spectrum is slightly redder.
It may be that quasar spectra from the rest-frame ultraviolet through the 
optical can be better represented by single power laws than suggested by
Francis et al. (1991), but this should be investigated more thoroughly
by examining spectrophotometry of individual objects with
wavelength coverage spanning from Lyman $\alpha$ through the Balmer lines.
The FBQS spectrum also has stronger
Lyman $\alpha$ and [O III] emission, and the C IV profile has a stronger
red wing than blue wing.  
The differences in the C IV profile and [O III] emission are also seen
in Figure 5, the comparison between the radio-loud and radio-quiet subsamples
of the FBQS.

Figure 5 fails to show a difference in Lyman $\alpha$, however.
Francis, Hooper, \& Impey (1993) created composite ultraviolet spectra of 
radio-quiet and radio-loud subsamples of the LBQS. These spectra differed at 
a marginally statistically significant level, in the sense that radio-loud 
quasars had stronger low-to-intermediate velocity emission, primarily from 
C IV and Lyman $\alpha$.  The FBQS may not be sampling sufficient numbers of
quasars, nor a sufficient range in radio-loudness.
The fraction of radio-quiet quasars containing the Lyman $\alpha$ emission 
line is only 25\% (out of $\sim$ 50 quasars with spectra that blue, 
cf. 39\% with log R* $<$ 1 in the FBQS as a whole), limiting
our statistics.  These highest redshift radio-quiet quasars also possess the 
largest log R* values of the radio-quiet class, and as radio loudness 
appears to be a more continuous property than once thought (White et al.
2000), it may not be unexpected that Lyman $\alpha$ is so similar in 
the radio-loud and radio-quiet subsamples of the FBQS.

That the FBQS composite has an unusually strong Lyman $\alpha$ cannot be
denied.  Cannonical values of C IV/Ly $\alpha$ lie within the range
$0.4--0.6$ (Baldwin et al. 1995), although most of these are drawn from
optically selected samples.  Using an estimate of 25\% contamination
from N V in the Lyman $\alpha$ blend, the FBQS composite spectrum shows
C IV/Ly $\alpha$ = 0.36, just outside the cannonical range.  
The C IV/(Ly $\alpha$ + N V) ratio from Francis et al. (1991) is 0.63, 
but the ratio in the updated LBQS composite is only 0.37, or 0.47 for 
C IV/Ly $\alpha$ after correcting for an estimated 25\% N V contamination.
That the Lyman $\alpha$ emission line is stronger in the FBQS 
than in the LBQS is not simply a matter of absorbed quasars.

This difference is probably related to the presence of strong extended
narrow-line regions/nebulosity seen preferentially in radio-loud objects
(e.g., Heckman et al. 1991).  Very large Ly $\alpha$/C IV ratios are seen in
high-redshift radio galaxies (e.g., McCarthy 1993) and the narrow-line regions
of Seyfert 2 galaxies (e.g., Ferland \& Osterbrock 1986).  Narrow-line
emission in general is stronger in radio-loud quasars (Boroson \& Green 1992).
These differences are tied into the so-called ``eigenvector 1'' relationships
of principal component analysis (e.g., Boroson \& Green 1992;
Francis et al. 1992; Brotherton et al. 1994; Wills et al. 1999), 
which places radio-loud quasars with strong narrow and intermediate emission 
lines at one extreme, and radio-quiet quasars and narrow-line Seyfert 1
galaxies at the other.   The current prevailing explanation for at least
a portion of these trends involves the variation in accretion rate and
covering fraction (e.g., Boroson \& Green 1992).

Sprayberry \& Foltz (1992) examined composite spectra of BAL quasar 
subsamples of the LBQS (see also Weymann et al. 1991).  
They concluded that the spectral shape of the
composite low-ionization BAL quasar composite spectrum was consistent
with that of the high-ionization BAL quasar composite spectrum if
reddened by an SMC-type extinction law with $E(B-V) \sim 0.1$.  This 
is rather similar to what we find for the FBQS, an entirely consistent
result given the small sample sizes and normalization uncertainties.
There remains the question of just how many low-ionization BAL quasars
were lost from the sample because of the magnitude limit; their existence
will probably make the true average low-ionization BAL quasar more reddened 
than this.  

\section{Summary}

We have created a composite quasar spectrum of the total non-BAL FBQS
sample and additional subsamples of particular interest, including
radio-quiet, radio-loud, and high and low ionization BAL quasars.  The
composite spectra are publically available via the FIRST Survey web
page\footnote{http://sundog.stsci.edu.}.  
The process used to create the composites
is identical to that used in creating the composite spectra of the LBQS 
(Francis et al.\ 1991; S. Morris 1999, private communication), to which we have
compared our FBQS composite spectra.  
While as robust as possible, there are unavoidable uncertainties inherent in
the procedure of creating composite spectra, 
and these uncertainties should be
considered when using such spectra for particular applications.

\acknowledgments

We thank Simon Morris, Kirk Korista, and Paul Francis for making available 
their software and LBQS composite spectra, as well as for additional 
commentary on this manuscript and the originally published LBQS 
composite spectrum.  We thank everyone who has contributed or is continuing 
to contribute to the FIRST Bright Quasar Survey.
The success of the FIRST survey is in large measure due to the generous
support of a number of organizations.  In particular, we acknowledge
support from the NRAO, the NSF (grants AST-98-02791 and AST-98-02732),
the Institute of Geophysics and Planetary Physics (operated under the
auspices of the U.S. Department of Energy by Lawrence Livermore
National Laboratory under contract No.~W-7405-Eng-48), the Space
Telescope Science Institute, NATO, the National Geographic Society
(grant NGS No.~5393-094), Columbia University, and Sun Microsystems.


\clearpage
\begin{deluxetable}{lcccccc}
\tablecaption{Line Strengths}
\tablewidth{0pt}
\tablehead{Emission Feature& Rest $\lambda$ & Start\tablenotemark{a} & End\tablenotemark{a} & Intensity\tablenotemark{b} & $\sigma$\tablenotemark{c} & EW$_{rest}$ \nl
& (\AA) & (\AA) & (\AA) & & & (\AA)}
\startdata
Ly$\beta$ + O VI & 1026 \& 1034 & 1018 & 1054 & 11.6 & \nodata & 11 \nl
Ly$\alpha$ + N V & 1216 \& 1240 & 1186 & 1286 & 100 & 30 & 87 \nl
O I & 1302 & 1288 & 1325 & 2.4 & \nodata & 2.3 \nl
C II & 1335 & 1325 & 1354 & 0.7 & \nodata & 0.7 \nl
Si IV + O IV] & 1400 & 1353 & 1454 & 6.8 & 1.1 & 7.3 \nl
C IV & 1549 & 1452 & 1602 & 27 & 11 & 33 \nl
He II + O III] & 1640 \& 1663 & 1602 & 1700 & 5.2 & 0.4 & 7.0 \nl
Al III + C III] & 1858 \& 1909 &  1828 & 1976 & 10 & 2.9 & 17 \nl
2000 feature &\nodata & 1985 & 2018 & 0.2 & \nodata & 0.4 \nl
2080 feature &\nodata & 2035 & 2125 & 1.2 & \nodata & 2.2 \nl
C II] & 2326 & 2242 & 2388 & 1.4 & \nodata & 3.0 \nl
[Ne IV] & 2423 & 2386 & 2464 & 0.4 & \nodata & 0.8 \nl
Mg II & 2798 & 2650 & 2916 & 13 & 4.4 & 34 \nl
2970 feature &\nodata & 2908 & 3026 & 1.3 & \nodata & 3.7 \nl
3130 feature &\nodata & 3100 & 3156 & 0.2 & \nodata & 0.8 \nl
3200 feature &\nodata & 3156 & 3236 & 0.3 & \nodata & 1.1 \nl
[Ne V] & 3346 & 3324 & 3372 & 0.1 & \nodata & 0.3 \nl
[Ne V] & 3426 & 3392 & 3452 & 0.4 & \nodata & 1.5 \nl
[O II] & 3727 & 3712 & 3742 & 0.6 & 0.6 & 2.6 \nl
[Ne III] + He I & 3869 \& 3889 & 3804 & 3934 & 1.1 & \nodata & 5.5 \nl
[Ne III] & 3968 & 3934 & 4012 & 0.4 & \nodata & 2.2 \nl
[S II] + H$\delta$ & 4068/4076 \& 4102 & 4044 & 4148 & 0.8 & \nodata & 4.4 \nl
H$\gamma$ + [O III]  & 4340 \& 4363 & 4276 & 4405 & 2.8 & 0.1 & 12 \nl
H$\beta$ & 4861 & 4704 & 5112 & 10 & 2.8 & 75 \nl
[O III] & 4959 & 4942 & 4976 & 1.2 & 1.3 & 7.5 \nl
[O III] & 5007 & 4986 & 5044 & 5.3 & 6.5 & 34 \nl
He I & 5876 & 5743 & 6015 & 1.7 & \nodata & 17 \nl
H$\alpha$ & 6563 & 6400 & 6800 & 31 & 12 & 300 \nl
Fe II COMPONENTS: & & & & & & \nl
1 & \nodata & 1610 & 2210 & \nodata & \nodata & \nodata \nl
2 & \nodata & 2210 & 2730 & 27 & 5.5 & \nodata  \nl
3\tablenotemark{d} & \nodata & 2960 & 4040 & 35 & 2.2 & \nodata  \nl
4 & \nodata & 4340 & 4830 & 3.6 & 0.8 & \nodata  \nl
5 & \nodata & 5050 & 5520 & 2.0 & 1.1 & \nodata  \nl
\enddata
\tablenotetext{a}{The line flux is integrated between these limits, which are the same as used by Francis et al. (1991).}
\tablenotetext{b}{Percent of flux of Ly$\alpha$ + N V blend.}
\tablenotetext{c}{Standard deviation, under the assumption that the line variation is independent of that of the continuum.}
\tablenotetext{d}{This component includes the Balmer continuum.}
\end{deluxetable}

\clearpage

\begin{deluxetable}{lcccc}
\tablecaption{Summary of Samples \& Select Composite Properties}
\tablewidth{0pt}
\tablehead{{\bf Sample} (N) & Selection & $<z>$ & N$_{peak}$\tablenotemark{a} & $\alpha$\tablenotemark{b} \nl
Subsample (\%) & & & (\AA) &  }
\startdata
{\bf FBQS} (657) & FIRST, $E < 17.8 $ & 1.0 & $\sim$ 2900  & $-$0.46 \nl
Radio-loud (61\%) & log R* $>$ 1  & 1.1 & $\sim$ 2800 & $-$0.50 \nl
Radio-quiet (39\%) & log R* $<$ 1 & 0.9 & $\sim$ 3200 & $-$0.42 \nl
&  & & \nl
{\bf LBQS} (1018) & Optical techniques& 1.3 & $\sim$ 2200 & $-$0.32 \nl
Francis et al. 1991 (68\%) &  $16.0 \leq m_{B_J} \leq 18.85$  &  1.3 & $\sim$ 2200  & $-$0.32 \nl
\enddata
\tablenotetext{a}{The approximate wavelength at which the histogram of the number of quasars contributing to the composite spectrum peaks.}
\tablenotetext{b}{The optical/ultraviolet power-law index fit between 1450 \AA\ and 5050 \AA, F$_{\nu} \propto \nu^{\alpha}$.}
\end{deluxetable}

\clearpage

\newpage
\begin{figure}
\psfig{file=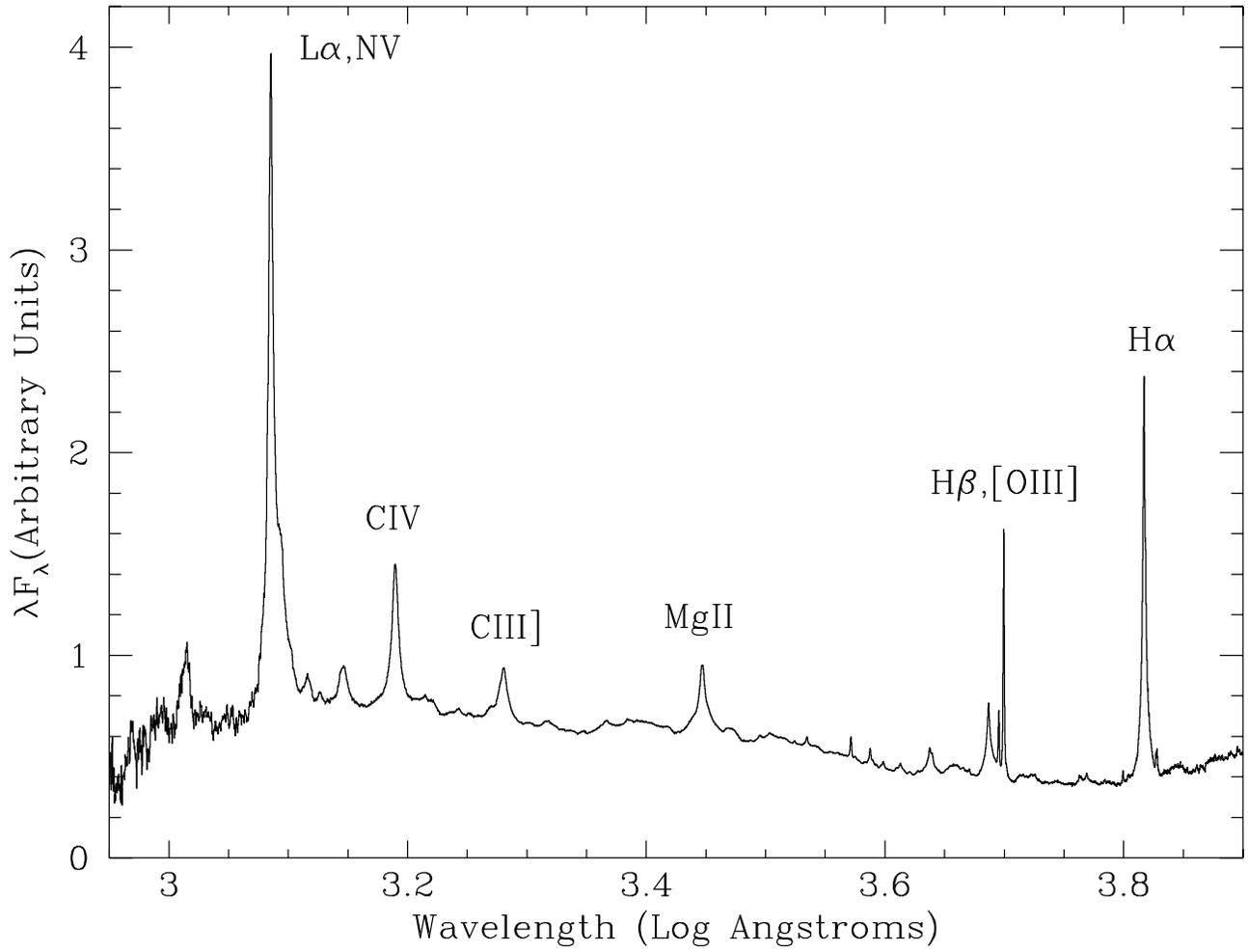,angle=-90,height=14cm}
\figcaption{FBQS composite spectrum plotted as $\lambda F_{\lambda}$ vs.\ the
logarithm of the rest-frame wavelength. Prominent emission features are marked.
The y-axis has been normalized such that the spectrum has values on order of 
unity.}
\end{figure}

\newpage
\begin{figure}
\psfig{file=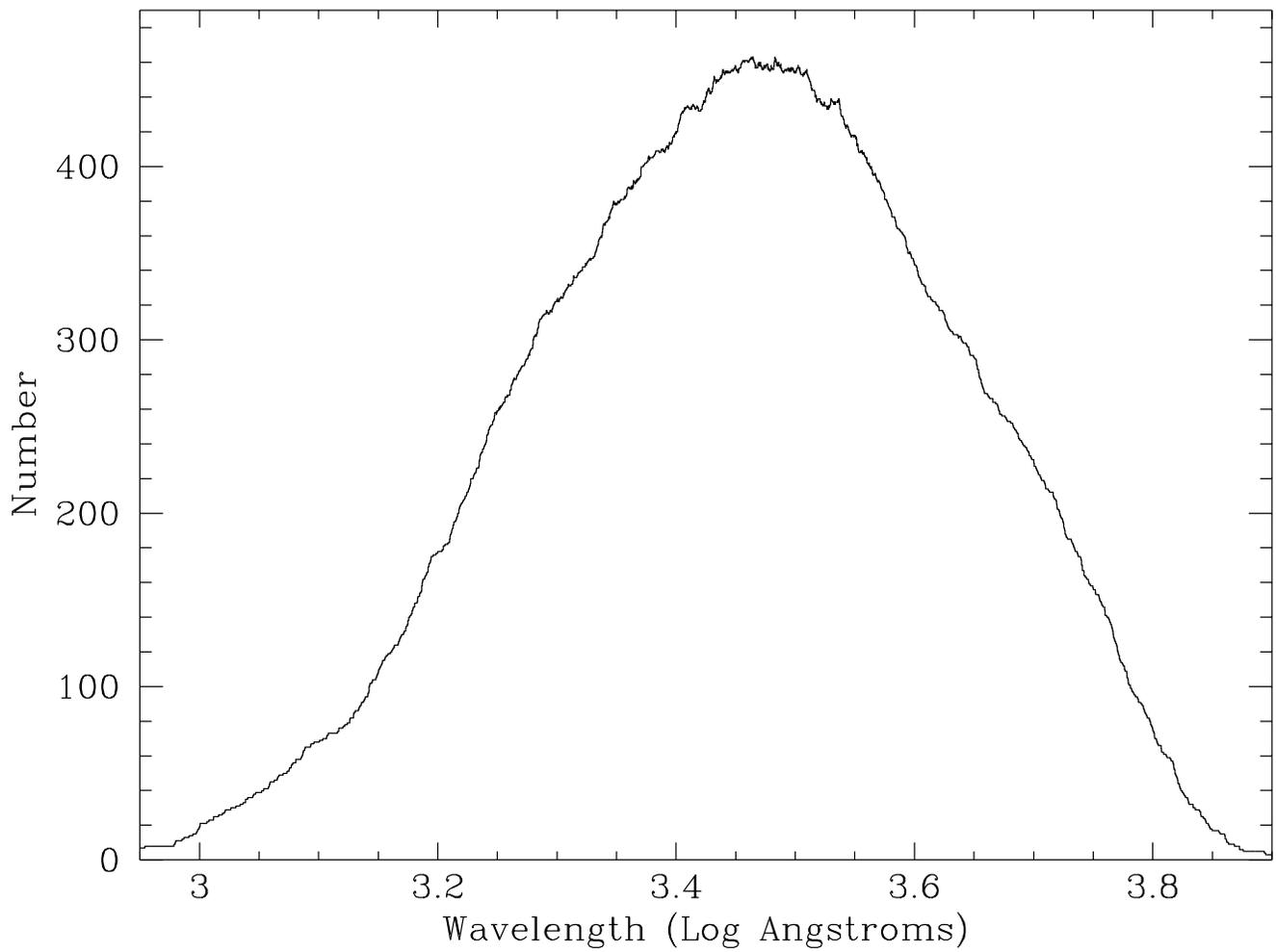,angle=-90,height=14cm}
\figcaption{Histogram of the number of quasars contributing at each wavelength
to the FBQS composite spectrum.}
\end{figure}

\newpage
\begin{figure}
\psfig{file=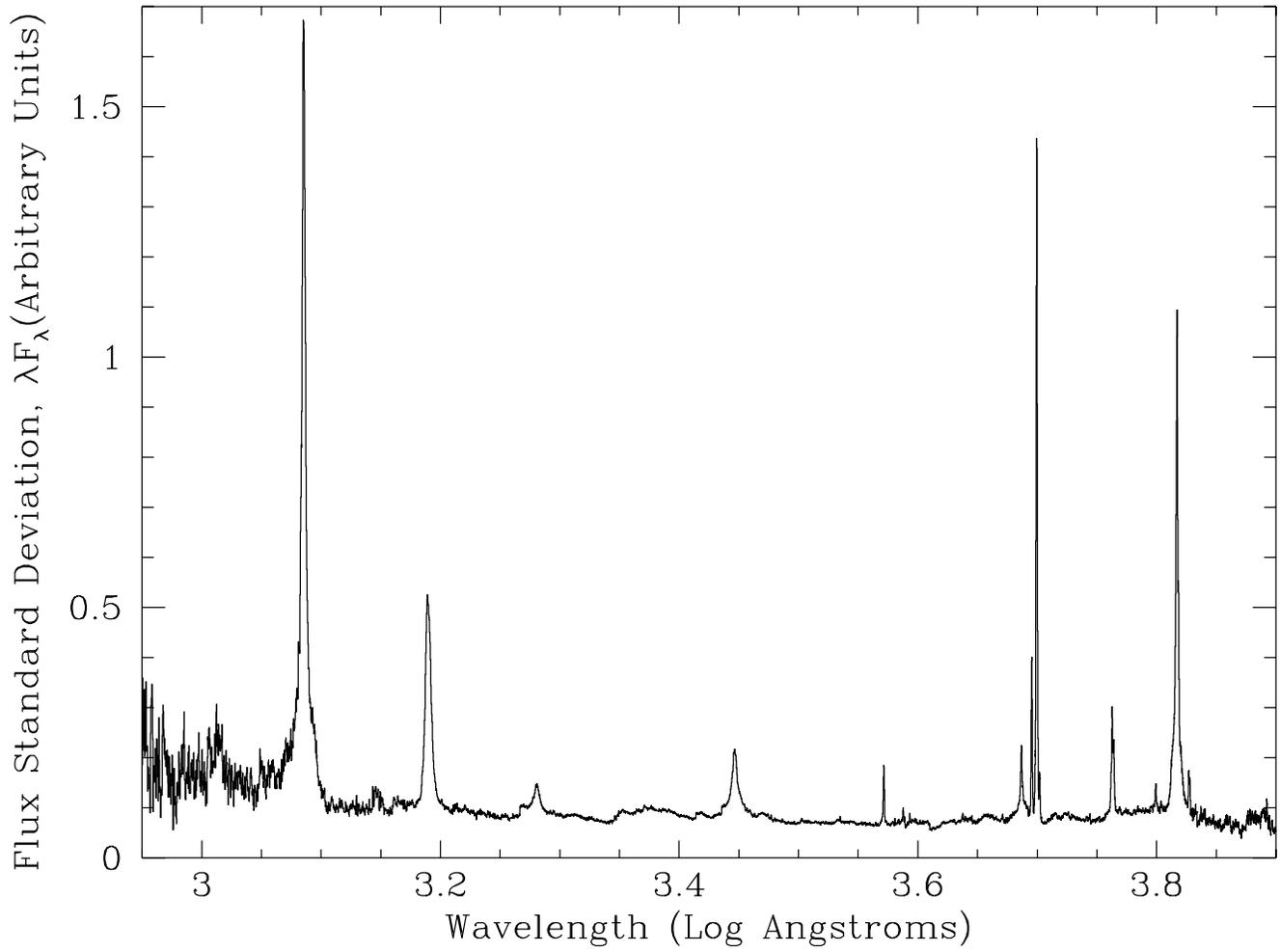,angle=-90,height=14cm}
\figcaption{Standard deviation in $\lambda F_{\lambda}$ relative to the
FBQS composite spectrum as a function of wavelength for the individual
spectra comprising the composite.  The same normalization constant has
been used as in Fig. 1.}
\end{figure}

\newpage
\begin{figure}
\psfig{file=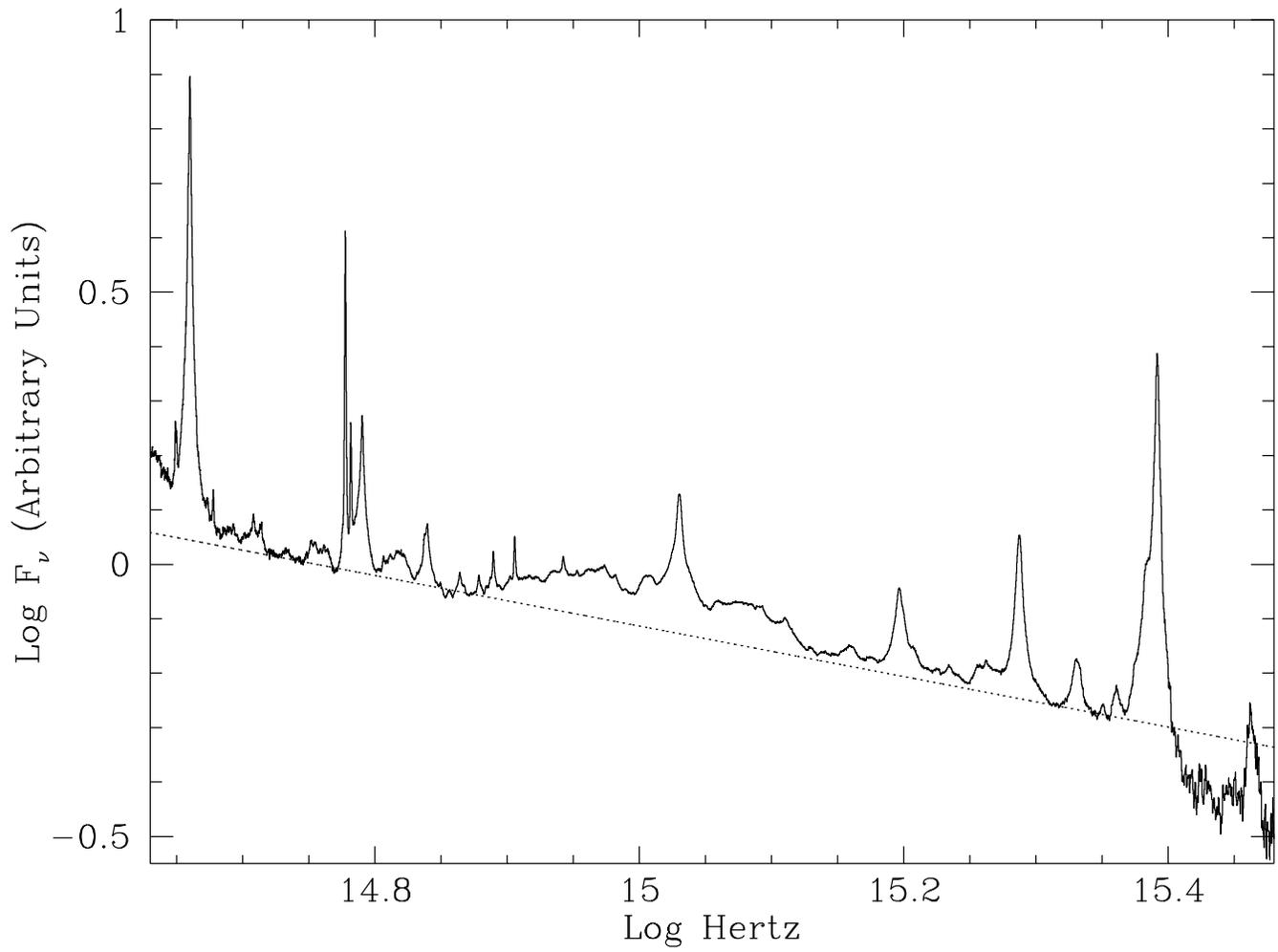,angle=-90,height=14cm}
\figcaption{Log(F$_{\nu}$)-log($\nu$) plot of the FBQS composite (solid line)
compared with a power law (dotted line) of index $\alpha$ = $-0.46$ (where
F$_{\nu}$ $\propto$ $\nu^{\alpha}$).}
\end{figure}

\newpage
\begin{figure}
\psfig{file=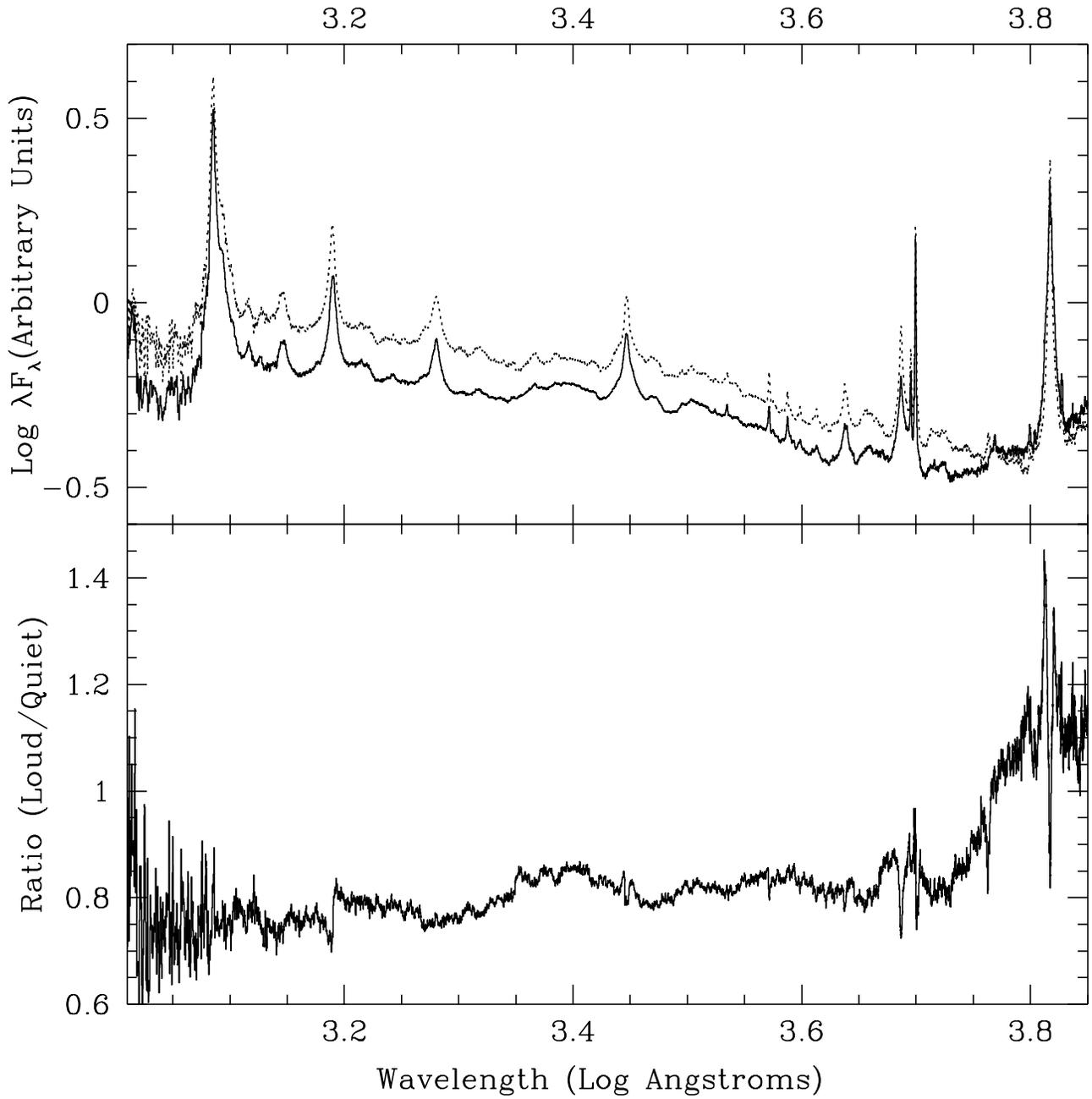,height=18cm}
\figcaption{Upper panel.  Comparison of mean spectra from the radio-loud 
(solid line) and radio-quiet (dotted line) subsamples of the FBQS.
Bottom panel.  Ratio of the radio-loud to radio-quiet composite spectrum.}
\end{figure}

\newpage
\begin{figure}
\psfig{file=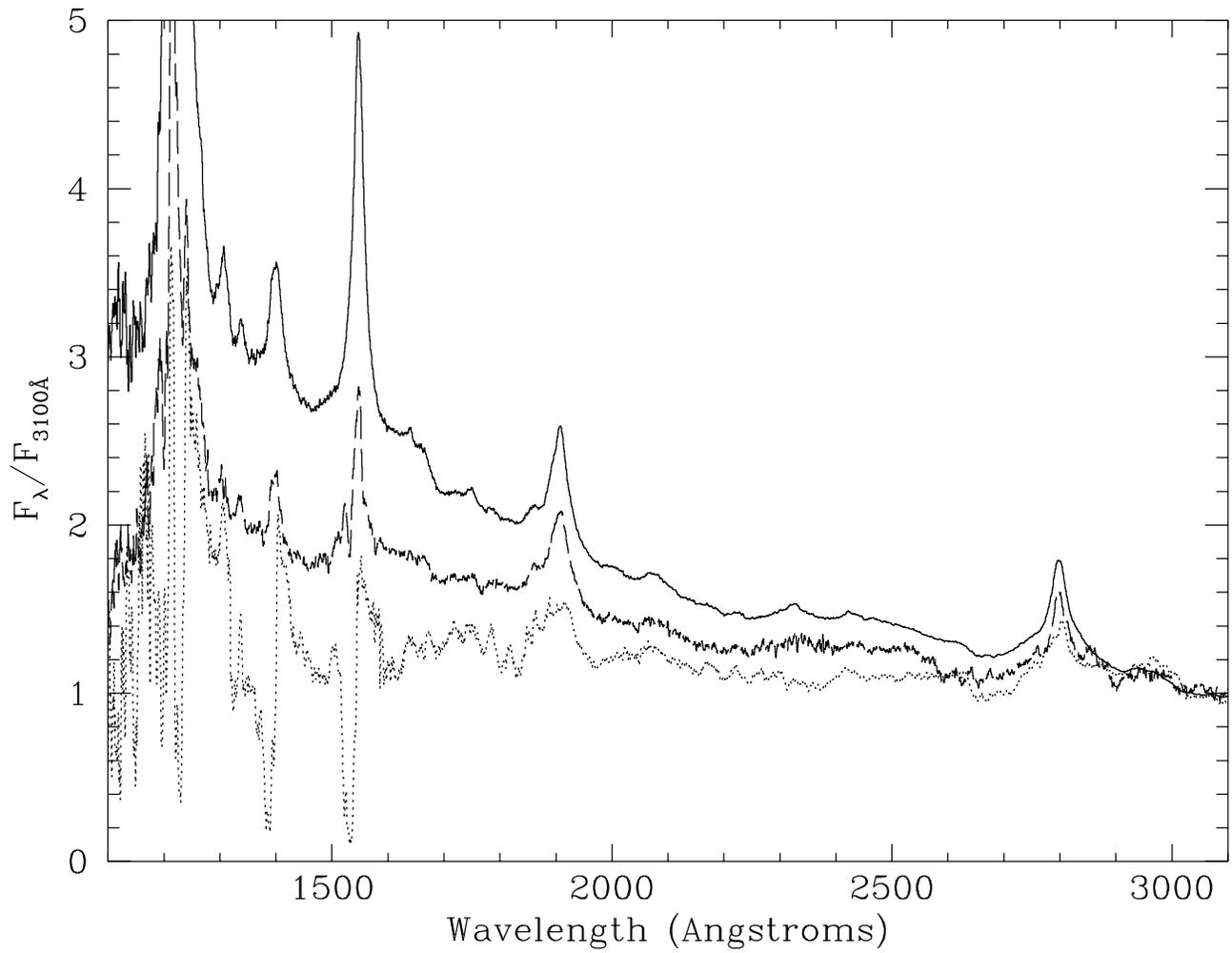,angle=-90,height=14cm}
\figcaption{Comparison of mean spectra from the FBQS (solid line) and subsamples
comprised of the high-ionization BAL quasars (dashed line) and low-ionization
BAL quasars (dotted line).  The spectra have been normalized to the flux at
3100 \AA.}
\end{figure}

\newpage
\begin{figure}
\psfig{file=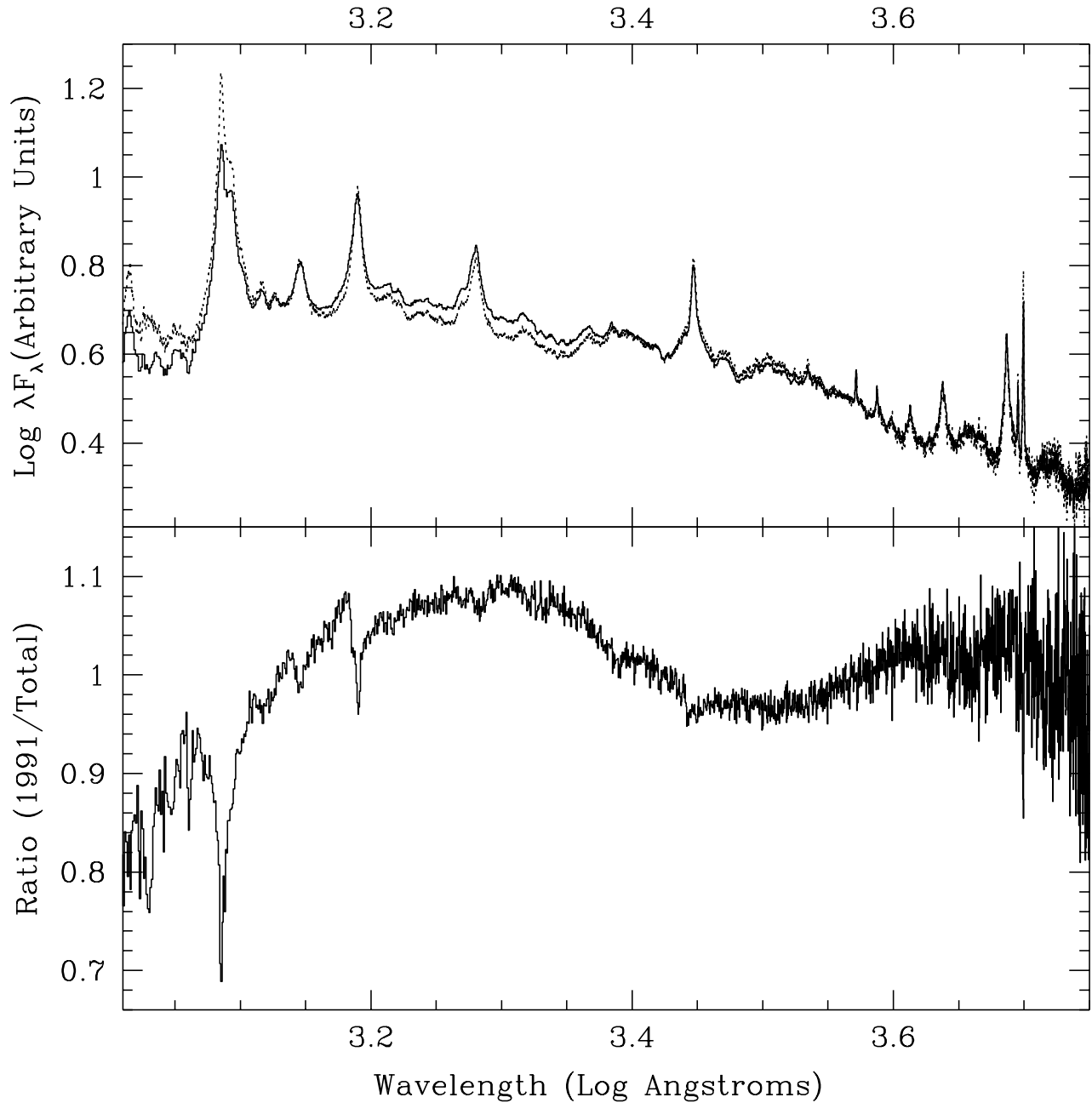,height=18cm}
\figcaption{Upper panel.  Comparison of mean spectra from the LBQS
as presented by Francis et al. (1991) (solid line) and updated with the
complete LBQS data set by Simon Morris (1999, private communication) 
(dotted line).
Bottom panel.  Ratio of published 1991 LBQS composite quasar spectrum to
that created with the total sample.}
\end{figure}

\newpage
\begin{figure}
\psfig{file=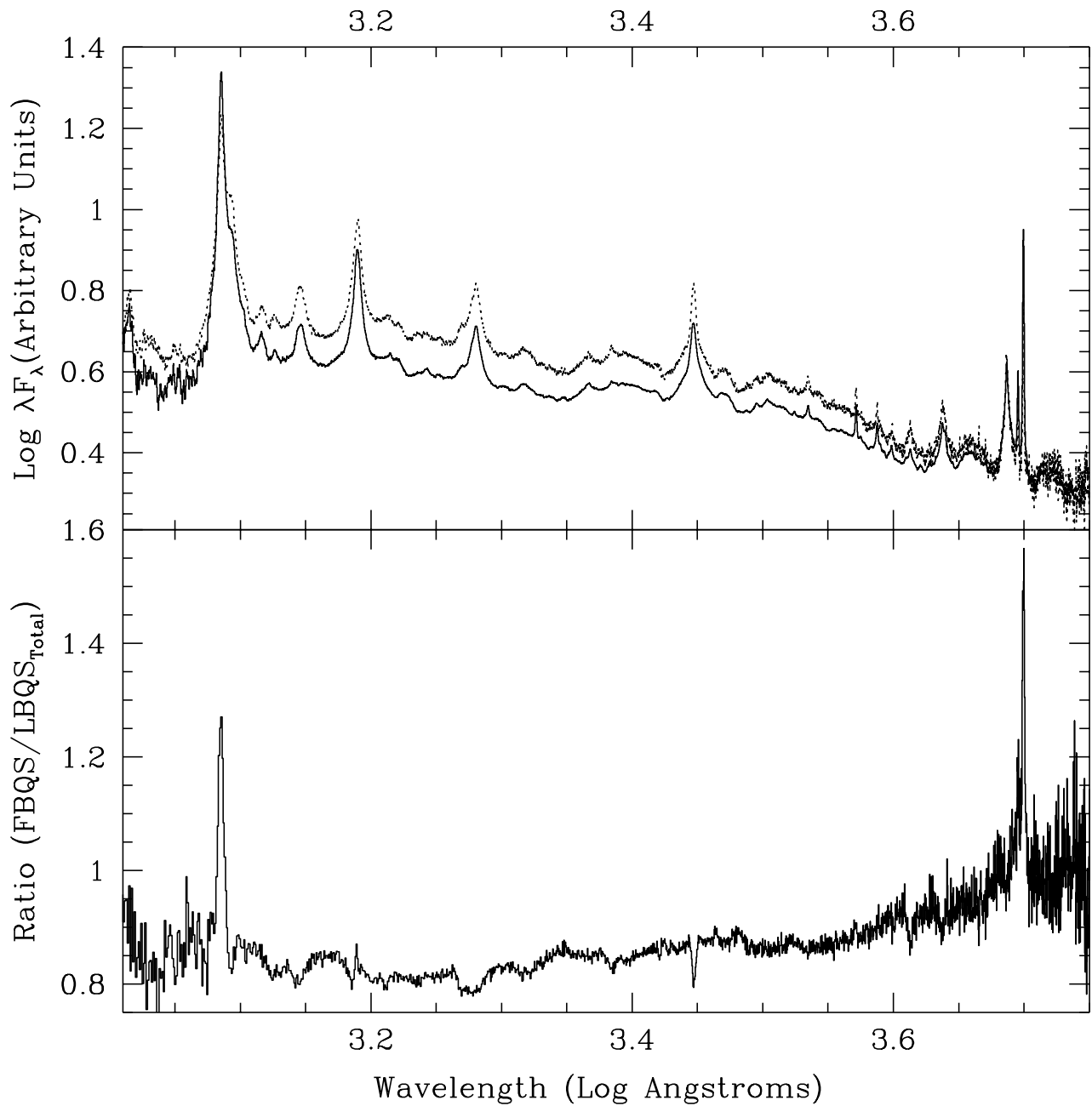,height=18cm}
\figcaption{Upper panel.  Comparison of mean spectra from the FBQS
(solid line) and the total LBQS (dotted line).
Bottom panel.  Ratio of the FBQS composite quasar spectrum to that of
the total LBQS sample.}
\end{figure}

\end{document}